\documentclass[twocolumn]{aastex631}

\graphicspath{{./}{figures/}}

\usepackage{threeparttable}
\usepackage{color}

\begin{document}

\title{Accelerated structural evolution of galaxies in a starbursting cluster at z=2.51}

\author{Can Xu}
\affiliation{School of Astronomy and Space Science, Nanjing University, Nanjing, Jiangsu 210093, China}
\affiliation{Key Laboratory of Modern Astronomy and Astrophysics, Nanjing University, Ministry of Education, Nanjing 210093, China}

\author[0000-0002-2504-2421]{Tao Wang}
\affiliation{School of Astronomy and Space Science, Nanjing University, Nanjing, Jiangsu 210093, China}
\affiliation{Key Laboratory of Modern Astronomy and Astrophysics, Nanjing University, Ministry of Education, Nanjing 210093, China}

\author{Qiusheng Gu}
\affiliation{School of Astronomy and Space Science, Nanjing University, Nanjing, Jiangsu 210093, China}
\affiliation{Key Laboratory of Modern Astronomy and Astrophysics, Nanjing University, Ministry of Education, Nanjing 210093, China}

\author{Anita Zanella}
\affiliation{Istituto Nazionale di Astrofisica (INAF), Vicolo dell’Osservatorio 5, I-35122 Padova, Italy}

\author{Ke Xu}
\affiliation{School of Astronomy and Space Science, Nanjing University, Nanjing, Jiangsu 210093, China}
\affiliation{Key Laboratory of Modern Astronomy and Astrophysics, Nanjing University, Ministry of Education, Nanjing 210093, China}

\author{Hanwen Sun}
\affiliation{School of Astronomy and Space Science, Nanjing University, Nanjing, Jiangsu 210093, China}
\affiliation{Key Laboratory of Modern Astronomy and Astrophysics, Nanjing University, Ministry of Education, Nanjing 210093, China}

\author{Veronica Strazzullo}
\affiliation{INAF – Osservatorio Astronomico di Trieste, Via Tiepolo 11, 34131 Trieste, Italy}
\affiliation{IFPU – Institute for Fundamental Physics of the Universe, Via Beirut 2, 34014 Trieste, Italy}

\author{Francesco Valentino}
\affiliation{Cosmic Dawn Center (DAWN), Copenhagen, Denmark}
\affiliation{Niels Bohr Institute, University of Copenhagen, Jagtvej 128, 2200 Copenhagen N, Denmark}

\author{Raphael Gobat}
\affiliation{Instituto de Física, Pontificia Universidad Católica de Valparaíso, Casilla 4059, Valparaíso, Chile}

\author{Emanuele Daddi}
\affiliation{Université Paris-Saclay, Université Paris Cité, CEA, CNRS, AIM, 91191, Gif-sur-Yvette, France}

\author{David Elbaz}
\affiliation{Université Paris-Saclay, Université Paris Cité, CEA, CNRS, AIM, 91191, Gif-sur-Yvette, France}

\author{Mengyuan Xiao}
\affiliation{Department of Astronomy, University of Geneva, Chemin Pegasi 51, 1290 Versoix, Switzerland}

\author{Shiying Lu}
\affiliation{School of Astronomy and Space Science, Nanjing University, Nanjing, Jiangsu 210093, China}
\affiliation{Key Laboratory of Modern Astronomy and Astrophysics, Nanjing University, Ministry of Education, Nanjing 210093, China}

\author{Luwenjia Zhou}
\affiliation{School of Astronomy and Space Science, Nanjing University, Nanjing, Jiangsu 210093, China}
\affiliation{Key Laboratory of Modern Astronomy and Astrophysics, Nanjing University, Ministry of Education, Nanjing 210093, China}

\correspondingauthor{Tao Wang}
\email{taowang@nju.edu.cn}

\begin{abstract}

Structural properties of cluster galaxies during their peak formation epoch, $z \sim 2-4$ provide key information on whether and how environment affects galaxy formation and evolution. 
Based on deep HST/WFC3 imaging towards the z=2.51 cluster, J1001, we explore environmental effects on the structure, color gradients, and stellar populations of a statistical sample of cluster SFGs. We find that the cluster SFGs are on average smaller than their field counterparts. This difference is most pronounced at the high-mass end ($M_{\star} > 10^{10.5} M_{\odot}$) with nearly all of them lying below the mass-size relation of field galaxies. The high-mass cluster SFGs are also generally old with a steep negative color gradient, indicating an early formation time likely associated with strong dissipative collapse.  For low-mass cluster SFGs, we unveil a population of compact galaxies with steep positive color gradients that are not seen in the field. This suggests that the low-mass compact cluster SFGs may have already experienced strong environmental effects, e.g., tidal/ram pressure stripping, in this young cluster. These results provide evidence on the environmental effects at work in the earliest formed clusters with different roles in the formation of low and high-mass galaxies.
\end{abstract}


\section{Introduction} 
\label{sec:intro}

The cosmic epoch of redshift $z \sim 2-4$ marks an important phase of  mass assembly and galaxy transformation for clusters of galaxies, at least for the most massive ones. This is first speculated from galaxy archaeology studies in the local Universe~\citep{thomas2010}, and has now been confirmed by the discovery of a significant population of starbursting (proto)clusters at these redshifts~\citep{miley2006,casey2015}. Unlike their local counterparts, these structures generally exhibit a much larger fraction of star-forming galaxies (SFGs) and starbursts~\citep{Wang2016}. Studying the physical properties of the member galaxies in these structures is essential to constrain the role of dense environment in the star formation and quenching of cluster galaxies.

During the last decade, extensive efforts have been made in probing the physical properties of SFGs in high-z clusters, including star-forming main sequence, mass-metallicity relation, gas content, and star formation efficiency~\citep{koyama2013,valentino2015,shimakawa2018,dannerbauer2017,noble2017,stach2017,coogan2018,hayashi2017,rudnick2017,wang2018}. However, many of these studies yield different and sometimes controversial conclusions, which are likely driven by the biases in selecting member galaxies as well as the various types or evolution stages of (proto)clusters they probed.
Therefore, in order to achieve a full understanding on the environmental effects on SFGs, a census of complete samples of SFGs in various types of (proto)clusters is essential, which is unfortunately quite difficult for most previous studies due to observational limitations.

In addition to the aforementioned physical properties, the structure/morphologies of cluster galaxies provide another important avenue to study their formation process. In particular, the structure/morphologies of the SFGs in clusters carry key information on the involved environmental mechanisms. Most environmental mechanisms, e.g., tidal and ram pressure stripping, and galaxy interactions, all leave imprints in the structure/morphologies of SFGs. So far, most studies have primarily focused on the size evolution of quiescent galaxies in clusters at intermediate redshifts~\citep{Matharu2019}, structural properties of representative samples of SFGs in high-z clusters or protoclusters is still poorly constrained.

A few earlier studies show that cluster SFGs are generally smaller than their field counterparts at low redshift ($z\lesssim 1$, ~\citealt{Weinmann2009,Maltby2010,FernandezLorenzo2013,Cebrian2014}), a sign that environmental effects such as ram pressure and tidal stripping may be at work~\citep{Boselli2006,Boselli2022}.  At $z \gtrsim 1-2$, however, the situation is less clear. A few recent works show that there is little environmental dependence on the sizes of SFGs at these redshifts ~\citep{afanasiev2022,strazzullo2023}, while some other works focusing on (proto)clusters at higher redshifts with more active star formation reveal that a significant fraction of massive cluster SFGs appear to be more compact than field galaxies~\citep{Wang2016,Perezmart2023}. This indicates that these starbursting (proto)clusters, in which their massive SFG members are going through a major phase of mass assembly, may represent the best laborotary to witness environmental effects at work for cluster SFGs.

In this paper, we focus on structual properties of star-forming member galaxies in the z=2.51 cluster J1001~\citep{Wang2016}, one of the most extreme cases of starbursting clusters or protoclusters. We extend our previous work on the same structure with newly obtained HST/WFC3 F125W and F160W (rest-frame optical) imaging towards a more complete sample of SFGs, while previous studies were based on shallow HST/WFC3 F110W imaging (rest-frame UV) on a biased sample (mainly CO-detected) of member galaxies. In addition to the structural properties, the multi-band HST/WFC3 imaging permits a census of the color profiles and stellar population properties of member galaxies, enabling probing the underlying physics on their structural/morphological differences from field galaxies.

The paper is organized as follows. In Section~\ref{sec:data}, we give a brief description of the data and sample selection. We detail the method used for the estimation of galaxy size and other physical parameters in Section~\ref{sec:method}. In Section~\ref{sec:results}, we present the mass-size relation, color profiles and average spectral energy distributions (SEDs) of the cluster SFGs and their comparison to field galaxies. We then discuss the implications and physical origins of the observed differences between cluster and field galaxies, and summarize our main findings in Section~\ref{sec:discussion}. 
Throughout the paper we adopt a cosmology with $\Omega_{\rm m}= 0.3$, $\Omega_{\Lambda}= 0.7$ and $\rm H_{0}  = 70 km s^{-1}  Mpc^{-1} $. Magnitudes are provided in the AB system ~\citep{Gunn1972}. We use \citet{bruzual2011} stellar population synthesis models and a Chabrier initial mass function (IMF,~\citealt{chabrier2003}).

\begin{figure*}[]
\centering
\includegraphics[scale=0.8]{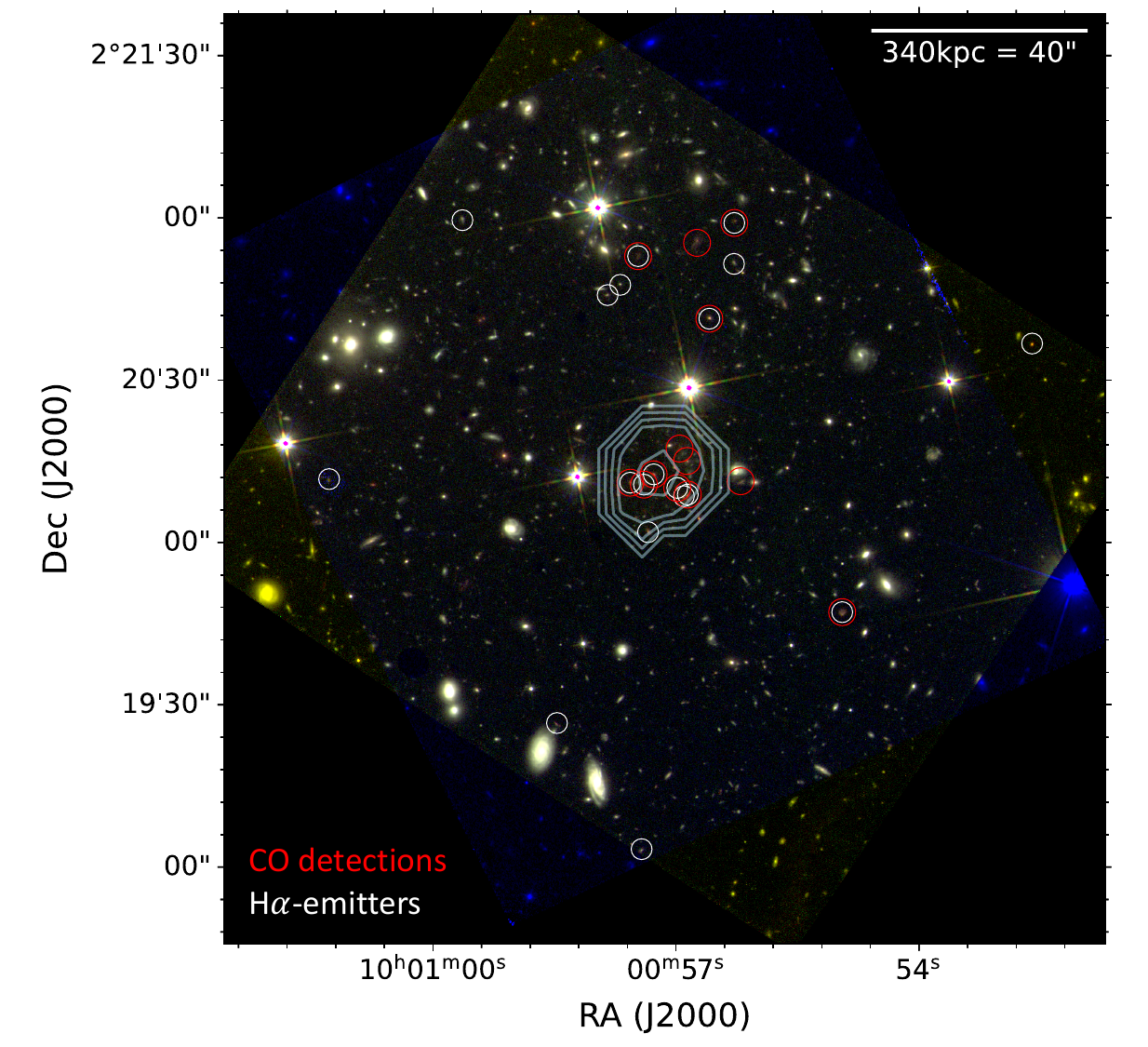}
\caption{RGB composite color image  of the $2.7\arcmin \times 2.7\arcmin$ field centered on the cluster J1001. The $R, G$, and $B$ channel of the image corresponds to the F160W, F125W, and F110W bands from HST/WFC3, respectively. Extended X-ray emission (0.5-2keV) are overlaid with grey contours. The positions of the 23 star-forming members are shown with circles, including 19 H$\alpha$ emitters(white) and 13 CO detections(red).  }
 \label{fig01-cluster}
\end{figure*}

\section{Data: member galaxies selection and HST imaging} 
\label{sec:data}

Our primary selection of the star-forming~(SF) members of J1001 is based on the deep narrow-band imaging with Subaru/MOIRCS~\citep{wang2018}, aiming to identify H$\alpha$ emitters at $z=2.49-2.52$ with the ``CO'' filter. We have detected 49 H$\alpha$ emitters with a dust-free SFR limit of $\sim~5 M_{\odot}$yr$^{-1}$~\citep{kennicutt1998}, which corresponds to lower mass limit (1$\sigma$)~of~$10^{9.2}M_{\odot}$ assuming the main sequence parametrization of~\citet{schreiber2015}. While this ensures that we are able to detect most of the SF members above this mass limit, some of the most obscured ones may still be missed. We hence complemented this sample of H$\alpha$ emitters with our previously confirmed cluster members based on CO(1-0)~\citep{Wang2016,wang2018} and CO(3-2)~\citep{xiao2022} observations.

The HST/WFC3 F125W and F160W imaging of J1001 is from Project 14750 (PI: T. Wang), which reaches 5$\sigma$ detection limit of $H_{\rm F160W} = 27.3$ for point sources. We further require HST/WFC3 coverage with a minimum of $H_{\rm F160W}$= 24.5, roughly the same cut as that in ~\citet{vanderWel2014}. By restricting H$\alpha$ emitters falling in the HST/WFC3 coverage, our final sample of SF members of J1001 includes 19 H$\alpha$ emitters, 4 of which are covered by our previous KMOS observations and have H$\alpha$ detections~\citep{Wang2016}. Some massive, and dusty obscured galaxies tend to have weak H$\alpha$ emission, which will be missed from the sample of H$\alpha$ emitters. By crossmatching this H$\alpha$-selected sample with the CO-detected members from previous work~\citep{Wang2016,wang2018,xiao2022}, we find 4 additional members detected in CO(1-0) or CO(3-2) lines~(Figure~\ref{fig01-cluster}), yielding a total of 23 SF members.

\begin{figure*}[]
\centering
\includegraphics[scale=1]{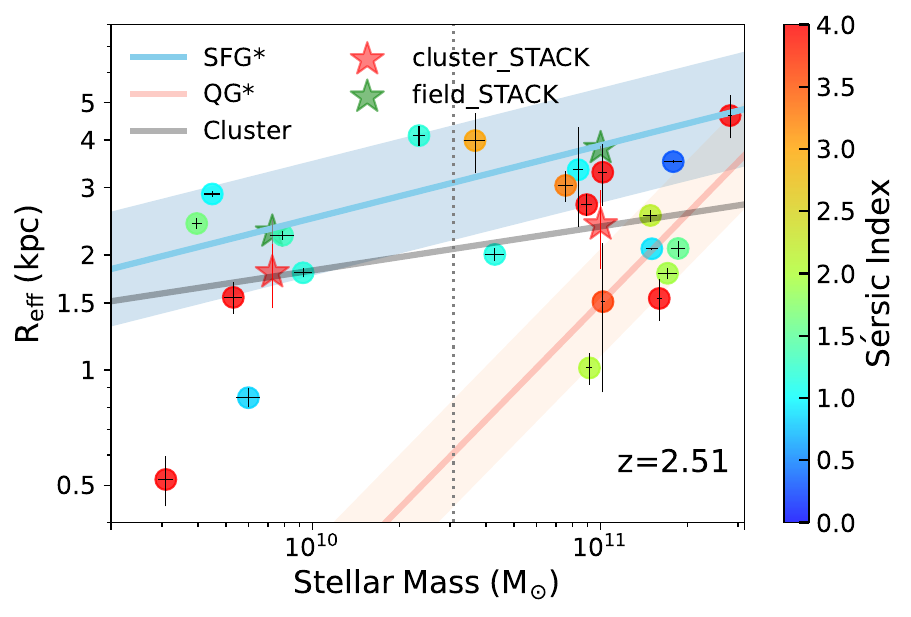}
\caption{Stellar mass-size distribution for cluster galaxies (fitting is shown by grey solid line). The relations with $1\sigma$ scatter for quiescent/star-forming field galaxies (red/blue solid line) are from~\citet[][]{vanderWel2014}. Cluster galaxies are plotted with circles. The colorbar represents the S{\'e}rsic index distribution of cluster galaxies. Grey vertical dotted line divide galaxies into high/low mass bins. Red/Green stars are for stacked cluster/field galaxies in low and high mass bins in our samples, respectively.}
 \label{fig02-M-R}
\end{figure*}

\section{Method} 
\label{sec:method}

\subsection{Derivation of structure properties and color profiles}

We use GALFIT ~\citep{Peng2011} to measure galaxy structural parameters. Point-spread function (PSF) of each band were created using TinyTim ~\citep{krist2011}. We adopted single S{\'e}rsic profiles~\citep{Sersic1963} consistent with \citet{vanderWel2014} for the fitting. For sources with nearby bright neighbors, we fit them simultaneously. Monte Carlo simulations are used to estimate the uncertainty of parameters reported by GALFIT. The effective radius ($R_{e}$) is defined as semi-major axis half-light radius from fitting $\rm F160W$ images. Considering the wavelength dependence of $R_{e}$, we applied the same correction as \citet{vanderWel2014} to get final $R_{e}$ estimates at rest-frame $\mathrm{5000\AA}$. 
We compared the fitting results with using PSFs from stacking of bright stars in the field, yielding consistent results (with typical difference in $R_{\rm e}$ less than $10\%$). We summarize our fitting results on the cluster galaxy samples in Appendix~\ref{figA01:highmass-extend}.

We measure surface brightness profiles and radial color profiles by Photutils~\citep{photutils1.6.0}\footnote{\url{https://photutils.readthedocs.io}} for stacking and individual galaxies. In the stacking procedure, the light-weighted center of each galaxy is shifted to the physical cut-out image center. We rescale the image onto a common grid and take the median pixel value as the flux of the stacked image at each position. Then we correct the different PSFs in the F125W and F160W bands by convolving with a Gaussian kernel to broaden the F125W-band images to match the angular resolution of the F160W-band images.

\begin{figure*}
\includegraphics[scale=0.40]{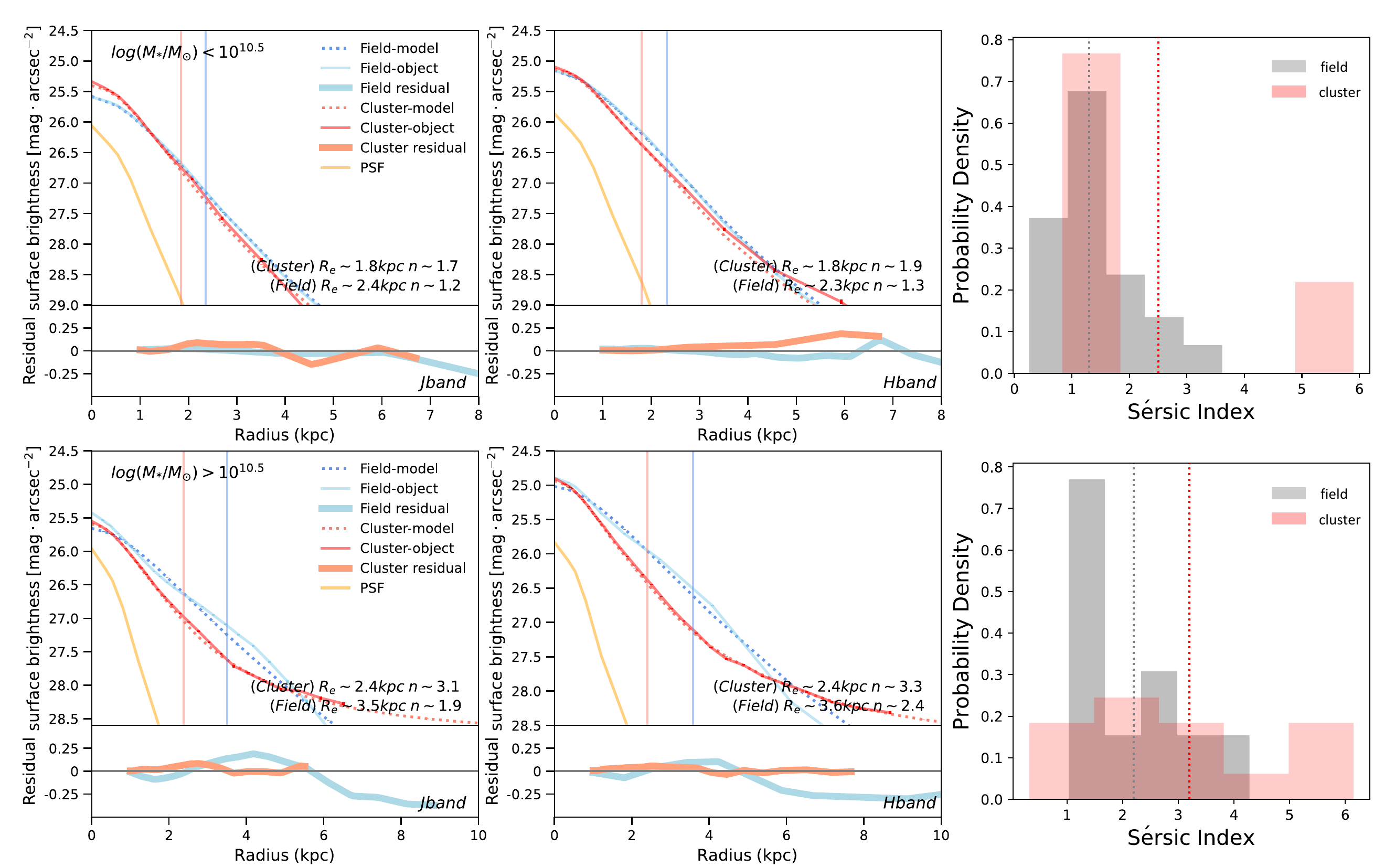}
\caption{Surface brightness profiles of cluster and field galaxies and their S{\'e}rsic index distributions. The top panels are shown for low-mass galaxies while the bottom panels are for high-mass galaxies. The surface brightness profiles at F125W and F160W are shown in the left and middle, respectively. The distribution of S{\'e}rsic index for cluster(red)/field(grey) galaxies derived from F160W band is shown on the right.} 

 \label{fig03-profile}
\end{figure*}

\subsection{Derivation of stellar masses}
\label{sec:method-masses}
We derive the stellar masses of our cluster galaxies by fitting their multiwavelength photometry from U-band to IRAC ch4 using the SED-fitting code BAGPIPES ~\citep[][]{carnall2018} with a MultiNest sampling algorithm. We use the stellar population synthesis models from \citet{bruzual2011} with a delayed exponential declining star formation history and the \citet{calzetti2000} extinction law. The nebular emission is constructed following the methodology of~\citet{byler2017}. We first run SED fitting with free stellar metallicity and get the initial stellar masses. Based on the mass-metallicity relation~\citep{kashino2022}, we estimated the prior range for metallicity in the new fitting. We repeated the SED fitting using the metallicity priors and obtained the final estimates for stellar masses, star formation rates, dust attenuation, and mass-weighted ages. We convert IMF from Kroupa~\citep{kroupa2002} to Chabrier by dividing stellar masses by a factor of 1.06.

\section{Results} 
\label{sec:results}


\subsection{Mass-Size Relation of cluster galaxies}
\label{sec:msr}

We show the mass-size relation for the cluster galaxies measured in the F160W band in Figure~\ref{fig02-M-R}. The relation for field galaxies is from~\citet{vanderWel2014}. Cluster SFGs appear to be systematically smaller compared to their field counterparts. This is more pronounced at the high-mass end ($M_{\ast} > 10^{10.5} M_{\odot}$), where half of the galaxies lie below the field relation considering the scatter. Galaxies at the low-mass end exhibits a large scatter, but on average, cluster SFGs are smaller~(~0.1dex) than their field counterparts, as also supported by the stacking results.

For a detailed cluster-field comparison, we consider a mass- and redshift-matched field sample based on the CANDELS/GOODS-South data\footnote{ Data were obtained from the Mikulski Archive for Space Telescopes (MAST) at the Space Telescope Science Institute; see \citet{candelsdata}. The specific observations analyzed can be accessed via \dataset[https://doi.org/10.17909/8gdf-dc47]{https://doi.org/10.17909/8gdf-dc47}.}~\citep{Grogin2011, Koekemoer2011}, which reach $H_{F160W}\sim 27$~(5$\sigma$) magnitude limits for point sources for imaging. Multiband photometry and stellar masses in GOODS-South are from \citet{Guo2013} and \citet{Santini2015}. SFGs at $2.25\le z \le 2.75$ are selected with the UVJ diagram~\citep{vanderWel2014}. We divide field/cluster galaxies into compact and extended subsamples depending on whether they are below or above the mass-size relation in the field/cluster, respectively. We then select three~(eight) field counterparts for each cluster galaxy in the same subsample within the stellar mass range of $\sim$0.2 dex in high-~(low-) mass bin. In total, 109 field SFGs are selected. As a sanity check, we stacked images of these field SFGs in two mass bins and derived their sizes with the same approach used for clusters, which are consistent with the field mass-size relation in \cite{vanderWel2014}.

As shown in the right panel of Figure~\ref{fig03-profile}, cluster SFGs exhibit larger S{\'e}rsic indexes than field galaxies. This is consistent with what we found in the stacked images, with best fitted S{\'e}rsic index of $3.3\pm0.3$ ($2.4\pm0.2$) and $1.9\pm0.4$ ($1.3\pm0.2$) for high-mass and low-mass cluster (field) SFGs. We perform T-tests to determine if there is a statistically significant difference between the average S{\'e}rsic index of cluster and field samples. The $p-$value is 0.044 ($\le0.05$) and 0.226 for the high- and low-mass galaxies. These results indicate more spheroidal morphologies for high-mass cluster SFGs, suggesting accelerated structural evolution in dense environments.

\begin{figure*}
\includegraphics[scale=0.65]{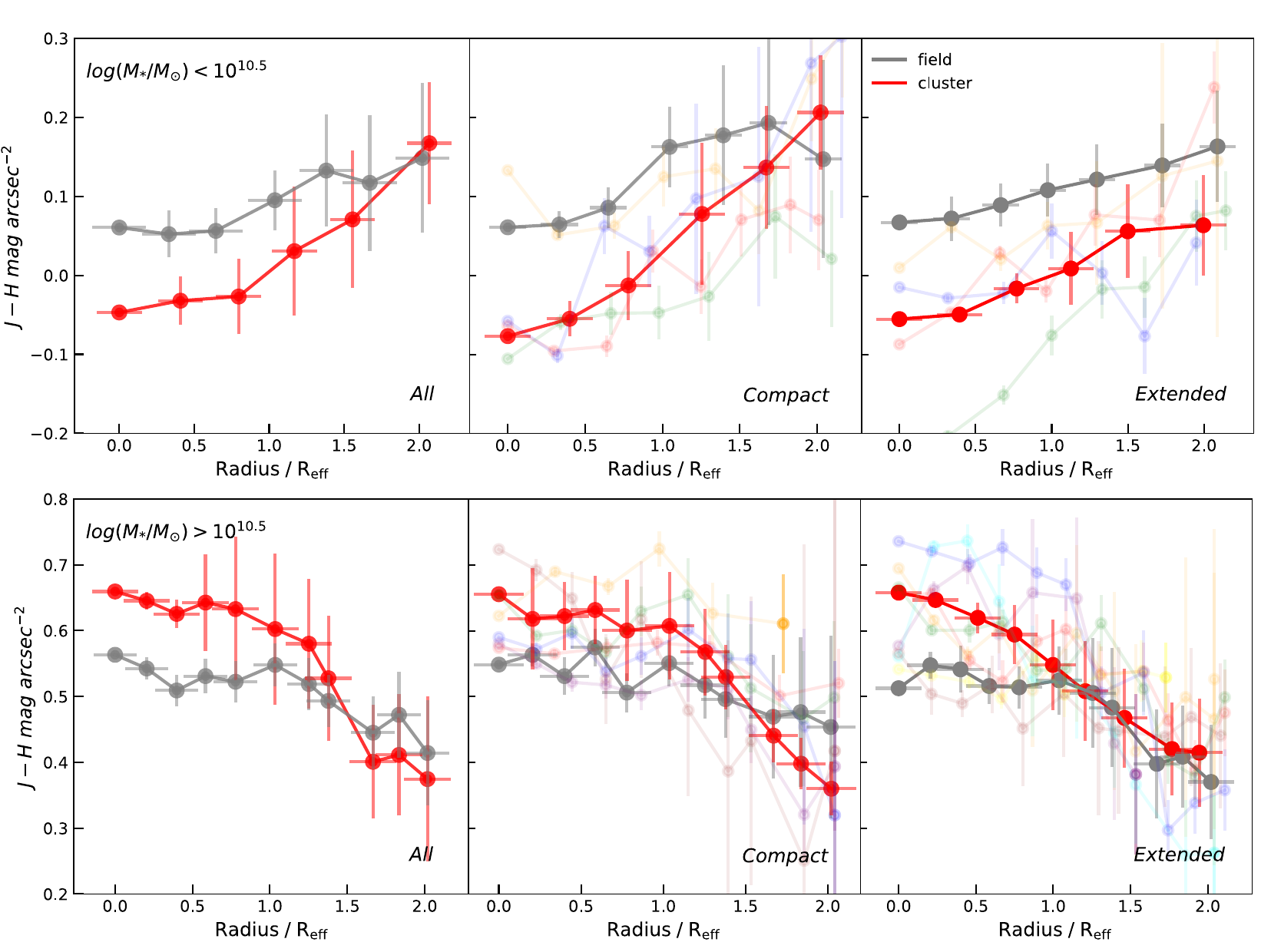}
\caption{Radial color profiles for stacked cluster(red)/field(grey) galaxies. Individual cluster galaxies are shown with rainbow-colored lines. Upper panels are for low-mass galaxies, and bottom panels are for massive galaxies. Left: stacked cluster/field galaxies. Middle/Right: divide cluster/field galaxies into compact/extended subsamples. }
 \label{fig04-color}
\end{figure*}

\subsection{The radial color profiles of cluster SFGs}
\label{sec:color}

We show the radial F125W $-$ F160W ($J - H$) color profiles for the cluster/field SFGs based on the stacked images in Figure~\ref{fig04-color}. In addition to the color profiles for the full sample, we also divide galaxies at both mass bins into compact and extended subsamples based on whether they are below or above the average mass-size relations shown in Figure~\ref{fig02-M-R}. We then derive the color profiles for each subsample (Figure~\ref{fig04-color}), for both stacked images and images of individual galaxies.
The maximum radial distance along the semimajor axis is set to $2R_{e}$. At the redshift of the cluster, $z = 2.51$, the $J - H$ color straddles the Balmer break, which is an excellent tracer of the average age of the stellar populations. Considering the effect of dust extinction, for example, assuming a Calzetti extinction curve ~\citep{calzetti2000}, $\Delta A_{V} = 0.3$ corresponds to only a change of $< 0.1$ dex in the $J - H$ color gradient at $z = 2.5$.

As shown in Figure~\ref{fig04-color}, distinct color gradients are revealed between field and cluster galaxies in both mass bins. The most striking feature is the steep positive color gradient for low-mass cluster galaxies, compared to the rather flat profile for field galaxies. For massive galaxies, on the other hand, a steeper negative color gradient is observed for cluster SFGs.

Dividing galaxies into compact and extended ones, we further show that compact low-mass galaxies exhibit steep positive color gradients. The extended low-mass cluster galaxies show similar color profiles as their field counterparts, but are on average bluer. For high-mass galaxies, the color profile is similar between compact and extended galaxies, both of which show a steeper negative gradient than their field counterparts. 
These results suggest that in addition to the structural properties, the cluster SFGs also behave differently in their color gradients, which may reflect their different star formation histories.

\begin{figure*}
\includegraphics[scale=0.18]{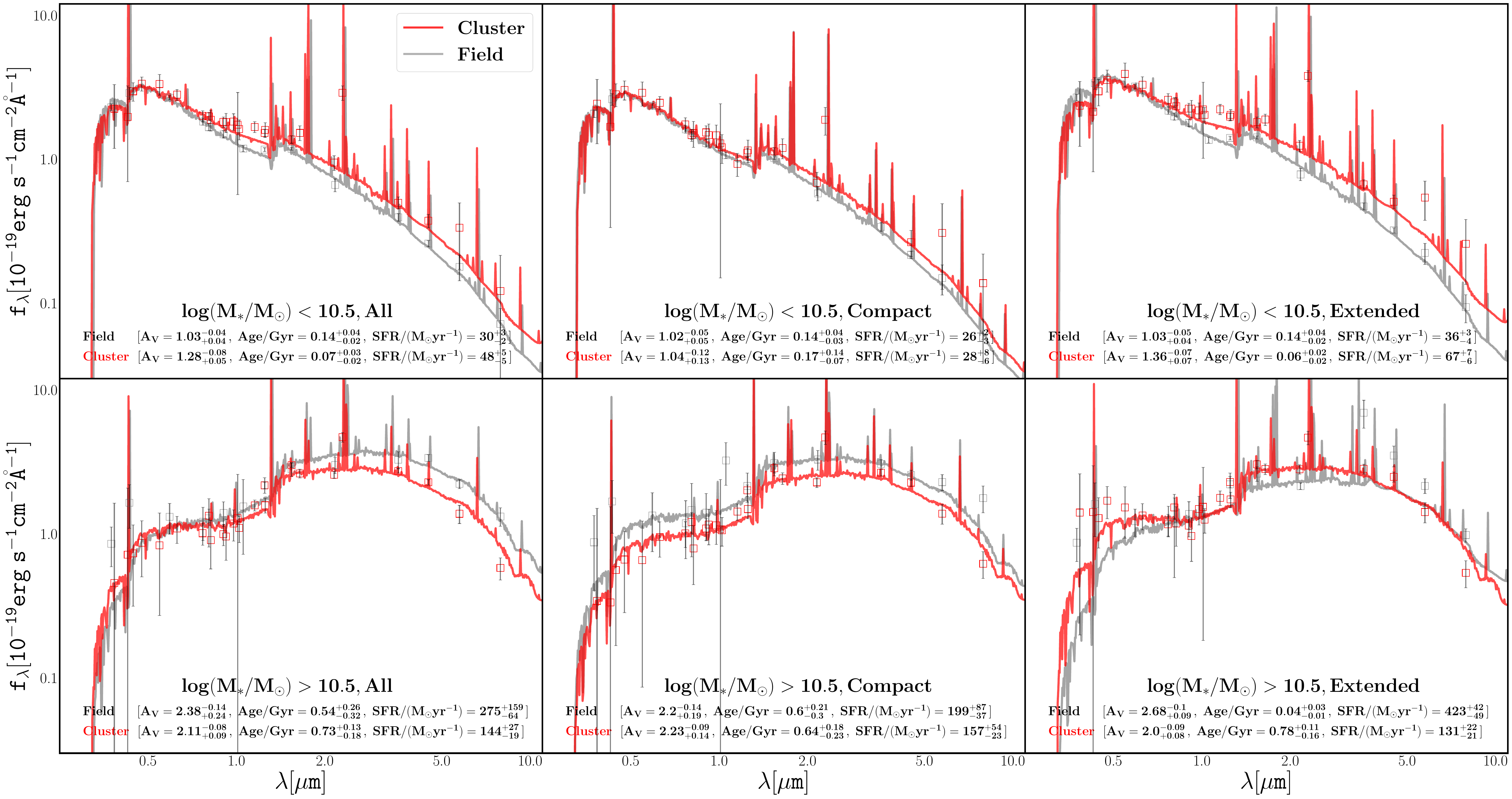}
\caption{Stacked SEDs for cluster and field galaxies. The upper panels are for low-mass galaxies, while the bottom panels are for high-mass galaxies. For both low- and high-mass galaxies, the stacked SEDs of all, compact and extended subsamples are shown in the left, middle, and right panels, respectively. The main physical parameters from the best-fitting SED are shown in each panel.}
 \label{fig05-SED}
\end{figure*}

\subsection{The average stellar population properties of cluster SFGs}
\label{sec:SED}

In order to examine general properties of the stellar populations of cluster SFGs, we derive the average SEDs, respectively for the low and high-mass subsamples, by computing the median flux densities with the Hodges-Lehmann estimator across U-band to IRAC ch4. The stacked SEDs for field galaxies are also derived with the same approach for comparison. We fit these stacked SEDs with BAGPIPES ~\citep[][]{carnall2018} using the same parameter setting as for SED fitting for individual galaxies introduced in \ref{sec:method-masses}. The stacked SEDs and our best-fitting results are shown in Figure~\ref{fig05-SED}.

For both low and high-mass galaxies, we find that the differences between cluster and field in their SEDs are mainly reflected in the extended population, while the SEDs of compact galaxies are rather similar. At the massive end, cluster galaxies are all relatively old, while the extended field galaxies are much younger, with a mass-weighted age of 0.04 Gyr, compared to 0.78 Gyr for extended cluster galaxies. The young age for the extended field galaxies is also clearly reflected in their weak Balmer break. For the extended low-mass galaxies, on the contrary, cluster galaxies are younger, consistent with their overall blue $J - H$ color profile~(Figure~\ref{fig04-color}). We argue that the lack of relatively young massive galaxies in the cluster indicates that massive cluster galaxies are generally formed earlier than their field counterparts. For low-mass cluster galaxies, on the other hand, the apparent age difference between compact and extended ones suggests that structural transformation is likely related to their accretion history onto the cluster.

\begin{figure*}
  \begin{center}
  \includegraphics[width=0.95\textwidth, angle=0 ]{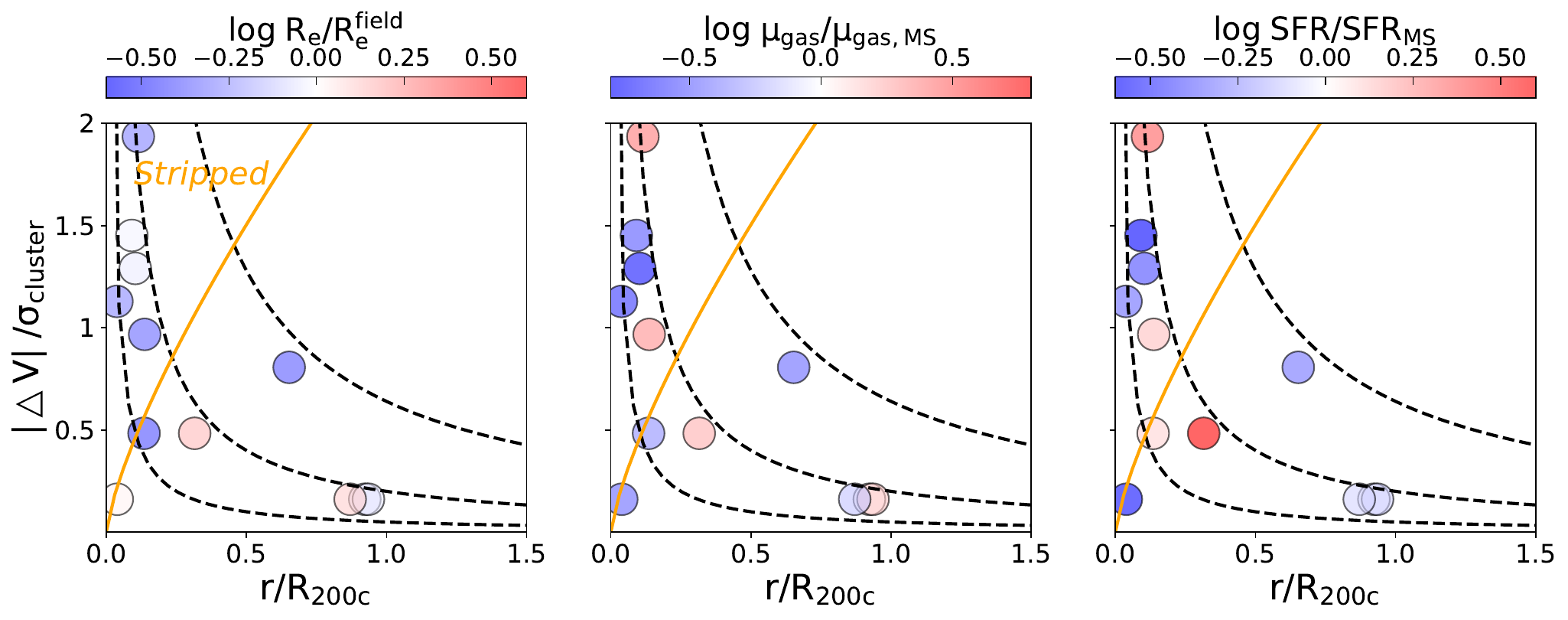} 
  \end{center}
  \caption{Phase-space diagram, which shows the clustercentric-radius dependence of $R_e$ (Left), gas fraction (Middle), and SFR (Right). Curves with $k_b$=0.05,0.2,0.64 are shown in the panels, with $\rm k_b=(r/R_{200c})\times (\left |\bigtriangleup V \right|\; /\sigma_{cluster})$. The upper-left stripped region contained by the orange line is defined following~\citet{Jaffe2015}. Cluster galaxies are plotted with circles and color-coded by their $R_e$ relative to the field mass-size relation,  their gas fractions and SFR are normalized by field main-sequence galaxies~\citep{Tacconi:2018}.  } 
  \label{fig06phase}
\end{figure*}

\subsection{Relation between galaxy sizes of cluster SFGs and their clustercentric radius}
\label{sec:SED}

Motivated by previous findings on a clustercentric-radius dependence of gas content in J1001~\citep{wang2018}, here we also explore whether and how the galaxy sizes change as a function of clustercentric radius.  We constructed line-of-sight velocity versus projected position phase-space diagrams~(Figure~\ref{fig06phase}) of the spectroscopically comfirmed member galaxies with the same approach as in~\citet{wang2018}. The phase-phase diagram allows us to infer the accretion history of these galaxies, where the parameter $\rm k_b=(r/R_{200c})\times (\left |\bigtriangleup V \right|\; /\sigma_{cluster})$ is roughly proportional to the time since infall. Galaxies with lower $k_b$ are more closely bound to the cluster. Since the cluster members in the outskirts lack HST/WFC3 coverage, our analysis is limited to those close to the cluster center within the virial radius $R_{200c}$, most of which are massive galaxies with $M_{\star} > 10^{10.5} M_{\odot}$. In addition to their sizes, we also show their gas content and SFRs (with the same data from \cite{wang2018}), enabling a comprehensive understanding on the clustercentric-radius dependence of the physical properties of cluster SFGs.

Figure~\ref{fig06phase} clearly shows that most of the compact cluster SFGs are preferentially located in the upper-left region that is expected to suffer from strong ram pressure stripping~\citep{Jaffe2015}. Moreover, those with lowest gas content and SFR (relative to similarly massive main-sequence SFGs) also tend to populate this region, providing further evidence that ram pressure stripping may be significantly affecting the structure, gas content and star formation properties in this distant cluster. However, not all compact galaxies in the ``stripped'' region exhibit low gas content and SFRs, which may reflects their different dynamical states and/or time delays between the changes in the stellar structure and ISM properties. We defer detailed discussions on this subject to a future work.

\section{Discussion and Conclusion}

\label{sec:discussion}

In this paper, based on a statistical and highly complete sample of SFGs (${\rm log}(M_{*}/M_{\odot})>10^{9.2}$) in the z=2.51 cluster J1001, we have revealed significant differences in the structure, color profiles, and stellar population properties between cluster and field galaxies, suggesting that the structural transformation and star formation quenching in cluster SFGs are accelerated. 
We summarize these findings as follows:
\begin{itemize}
    \item Cluster SFGs are on average smaller than their field counterparts~($\sim0.1$ dex), a difference that is most pronounced at the high-mass end ($M_{\star} > 10^{10.5}M_{\odot}$) with most of the high-mass SFGs~(86\%) lying below the mass-size relation of field galaxies (53\%, if considering the $1\sigma$ scatter). Cluster SFGs are also on average more spheroidal with higher S{\'e}rsic indexes at all masses.
    \item High-mass cluster SFGs exhibit steep negative color gradient in $J - H$ (rest-frame $U - B$), irrespective of their compactness. At low masses, they show an overall much bluer color than their field counterparts and the compact SFGs (lying below the mean mass-size relation of all cluster SFGs) exhibit steeper positive color gradients. 
    
    \item The stellar populations of high-mass cluster SFGs are relatively old irrespective of their compactness. For low-mass cluster SFGs, the compact galaxies are generally older than the extended ones.
 
\end{itemize}

The presence of strong mass dependence in the explored properties indicates different roles of the dense environment in the evolution of high- and low-mass systems.  We hence discuss the two populations separately.

\subsection{Early rapid formation of massive SFGs in starbursting clusters}
For the high-mass cluster SFGs, their overall smaller sizes (and higher S{\'e}rsic indexes) than field galaxies and old stellar populations indicate they are in a more advanced transition phase into quiescent galaxies.

Two effects may drive this accelerated transformation of the high-mass cluster SFGs. 
Firstly, the massive cluster SFGs likely formed most of their stars starting at earlier times than their field counterparts. So the size difference would naturally arise if the cluster SFGs are formed earlier when the Universe is denser. The similarity in the stellar populations between cluster and field compact SFGs suggests that the field compact SFGs may have similar formation history. 
Secondly, the overall steep negative gradients for both compact and extended cluster galaxies and the strong clustercentric dependence indicate that some other mechanisms are at work. We argue that such a steep negative color gradient indicates an inside-out growth scenario that most of the stars were likely formed during strong dissipative collapses in deep potential wells of galaxy cores~\citep{Tacchella2015}, which could be caused by intensive cold gas accretion in massive dark matter halos,  major gas-rich mergers, or ram pressure compression, all of which are facilitated in high-z (proto)clusters  close to the deep cluster potential~\citep{Treu2003}.

\subsection{Formation of low-mass compact SFGs with shrinking star-forming disks through gas stripping }
The low-mass, compact cluster SFGs are characterized by small sizes and steep positive color gradients, consistent with an outside-in quenching scenario~\citep{Kalita2022}. In comparison, the color gradients for the field SFGs are much less prominent. This suggests that the cluster SFGs have likely gone through ram pressure or tidal stripping events~\citep[][]{Gunn1972,Moore1999,Boselli2006,Boselli2022}, during which the gas in the outer disk was stripped off, and subsequently star formation in the outer disk was quenched. The major difference between tidal and ram pressure stripping is that tidal stripping can also strip the stellar component and form tidal tails. 
As shown in Figure~\ref{figA04:lowmass-compact}, most of these compact SFGs are rather isolated and do not have prominent tidal tails, indicating that either they have experienced tidal stripping at a much earlier time or the dominant mechanism is ram pressure stripping. In addition, compared to the extended cluster SFGs, the compact ones are older and have smaller SFRs, which suggests that they may have been accreted on the cluster earlier than the extended ones. Most likely, they have already survived a round-trip around the cluster core~\citep{cen2014}, where the ram pressure stripping effect is expected to be stronger. The extended ones, on the other hand, are likely accreted/formed very recently and have not been strongly influenced by the cluster environment.

~\\
\subsection{Comparison with previous works on J1001 and other high-z (proto)clusters}

Our results on the generally smaller sizes of massive galaxies are in good agreement with previous work on J1001 in \cite{Wang2016}, which studied mainly the structural properties of the high-mass galaxies in our sample based on HST/WFC3 F110W imaging (rest-frame UV). This study extends the size measurements to a larger sample and  lower stellar masses based on much deeper F125W and F160W (rest-frame optical) imaging.  More importantly, with both the F125W and F160W imaging, we could further explore their color profiles and build more accurate SEDs, allowing probing the underlying physics of their structural differences between field galaxies.

To put the findings on J1001 in the general context of galaxy formation in high-z clusters, here we compare our results on J1001 to other (proto)clusters at $z \gtrsim 2$. Unfortunately, due to difficulties in both member galaxy identification and obtaining high-resolution rest-frame optical imaging, there are not many structures at $z \gtrsim 2$ with detailed studies on the structural properties of a representative sample of SFG members. Among the very few (proto)clusters at $z \sim 2$ with detailed structural studies of a rather complete SFG sample, the spiderweb (proto)cluster has probably the best multiwavelength data and the least unbiased view on the mass-size relation on its member galaxies. As shown in ~\citet{Perezmart2023}, a population of massive compact SFGs do exist, which is similar to though less extreme than this study. On the other hand, such a population of massive compact SFGs appear to be absent in those more mature clusters at lower redshift~\citep{strazzullo2023}, which generally reveal similar mass-size relation as field galaxies.

In addition to the different types/evolutionary stages of structures, another potential factor that may cause different results on the mass-size relation for SFGs in high-z (proto)clusters is the selection of SFG members. In particular, Figure~\ref{fig06phase} shows that many of the compact SFG members tend to have lower SFR and/or low gas content. As a result, their identification depends strongly on the depth of the observations, including optical-to-NIR spectroscopy, narrow-band imaging and CO observations. In this sense, the SFG members of J1001 represent probably one of the least biased SFGs samples among $z \sim 2$ (proto)clusters by combining deep narrow-band imaging, NIR spectroscopy and  CO observations. A less complete SFG member sample, which misses those less active SFGs, would tend to drop those most compact ones, yielding a mass-size relation more closer to the field.

Keeping the aforementioned potential biases in various studies in mind, we emphasize that J1001 and other similar starbursting (proto)clusters is in a rapid transition phase from protoclusters or young clusters to mature clusters, with the most defining feature of a large concentration of massive SFGs (and starbursts) in the center of the (proto)cluster. Their large stellar masses, short gas depletion time, and compact sizes all suggest that they will soon be quenched and transit to quiescent galaxies. These may explain why strong environmental effects have been  found for general member galaxies in J1001, which may be also expected in other similar structures~\citep{Oteo2018,Miller2018,daddi2021}. In many aspects (dark matter halo masses, galaxy densities, et al.), their properties are already quite similar to mature clusters, which may explain why strong environmental effects have been observed. Very likely, these structures will soon evolve of the first mature clusters formed in the Universe with already a dominant quiescent galaxy population at $z \sim 2$~\citep{gobat2013,strazzullo2023}. In the future, more detailed studies on a larger number of similar structures as J1001 may finally tell us how these first clusters and their member galaxies are assembled.

To summarize, based on a complete sample of SF members in the cluster J1001 at $z = 2.51$, we find systematic differences in sizes, S{\'e}rsic indexes, color gradients, and stellar population properties between cluster and field galaxies. Our results provide clear evidence of the environmental effects in these young clusters, which are likely at the end of mass assembly but still with active star formation. Specifically, for high-mass cluster SFGs, their systematically smaller sizes, older stellar populations, and steep negative gradients suggest an early formation time likely associated with strong dissipation collapse. For low-mass galaxies, a population of compact SFGs with steep positive color gradients indicates the prevalence of tidal and/or ram pressure stripping events in high-z clusters. Future high-resolution resolved studies on the spatial distribution of their stars and gas components are required to consolidate the dominating physical mechanisms.

\begin{acknowledgments}

This work is supported by the National Natural Science Foundation of China (Project No. 12173017, and Key Project No. 12141301), and the China Manned Space Project with No. CMS-CSST-2021-A07. 

\end{acknowledgments}

\appendix
\restartappendixnumbering
\section{GALFIT FITTING RESULTS}

\setcounter{figure}{0}

\begin{figure}[ht!]
\epsscale{1.2}
\plotone{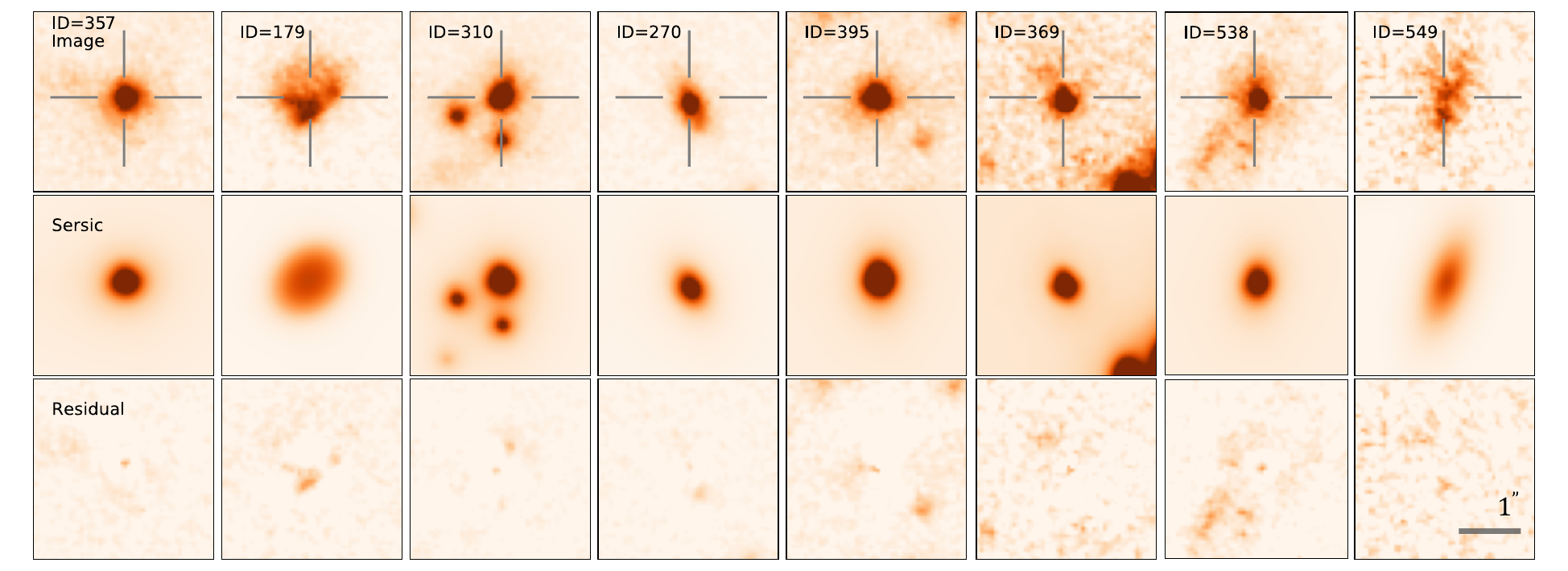}
\caption{Summary of GALFIT fitting of the 23 SFGs in J1001 with the three columns for HST/WFC3 H-band imaging, S{\'e}rsic models from GALFIT, and residuals between observations and models. This figure displays the 8 high-mass compact SFGs that are located above the grey solid line (Figure~\ref{fig02-M-R}) in decreasing stellar mass order. }
\label{figA01:highmass-extend}
\end{figure}

\begin{figure}[ht!]
\epsscale{1.2}
\plotone{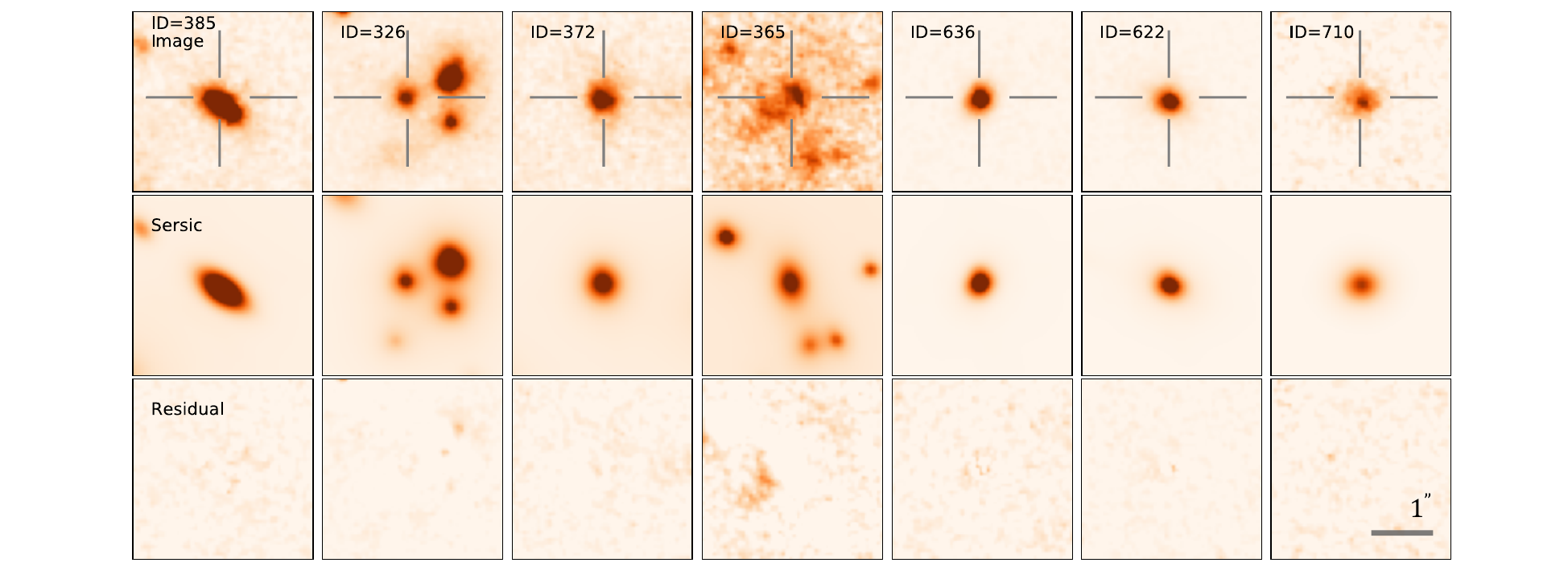}
\caption{Similar to Figure~\ref{figA01:highmass-extend}, but for the 7 high-mass compact SFGs.}
\label{figA02:highmass-compact}
\end{figure}

\begin{figure}[ht!]
\epsscale{1.2}
\plotone{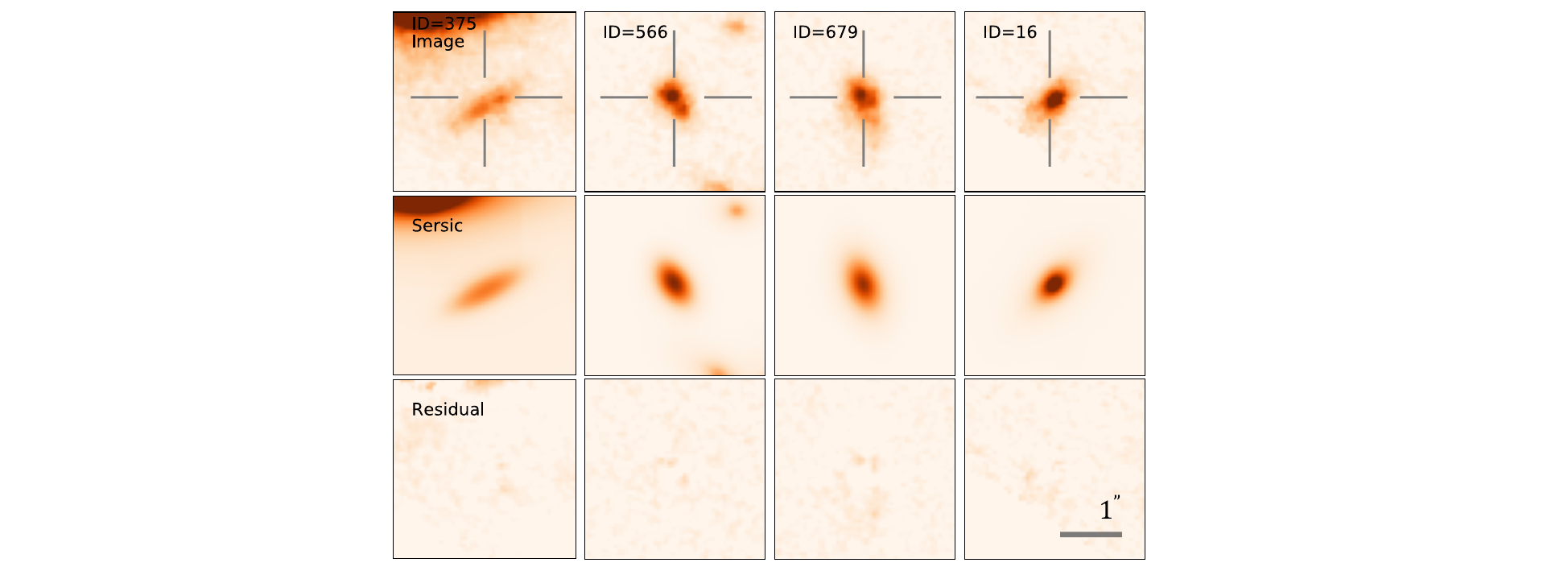}
\caption{Similar to Figure~\ref{figA01:highmass-extend}, but for the 4 low-mass extended SFGs.}
\label{figA03:lowmass-extend}
\end{figure}

\begin{figure}[ht!]
\epsscale{1.2}
\plotone{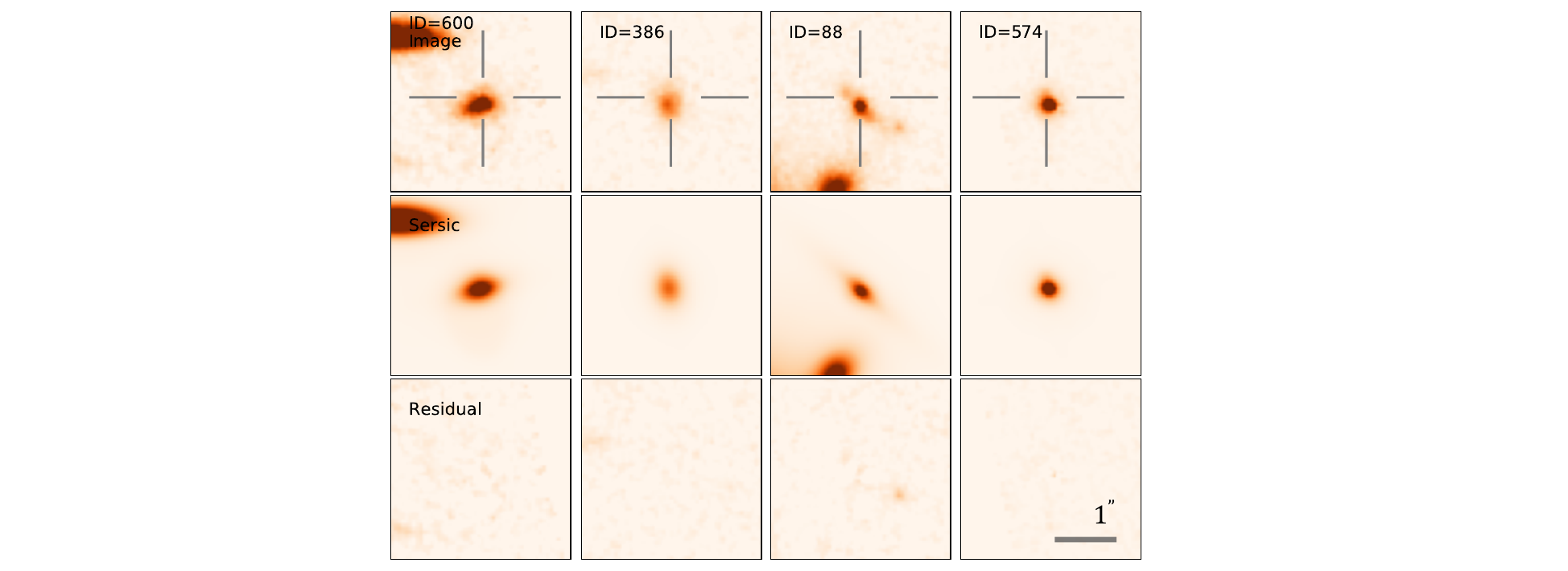}
\caption{Similar to Figure~\ref{figA01:highmass-extend}, but for the 4 low-mass compact SFGs.}
\label{figA04:lowmass-compact}
\end{figure}

The GALFIT fitting results~(Table~\ref{Tab}) of the 23 star-forming galaxies in J1001 are shown in Figure~\ref{figA01:highmass-extend}, Figure~\ref{figA02:highmass-compact}, Figure~\ref{figA03:lowmass-extend} and Figure~\ref{figA04:lowmass-compact}.

\begin{table*}[!ht]
\caption{ Confirmed Star-forming Cluster Members }\label{Tab}
\centering
\setlength{\tabcolsep}{1.5 mm}{
\begin{threeparttable} 
\begin{tabular}{c|c|c|c|c|c|c}
\hline\hline 
ID & RA & DEC  & 
Detections & log ($M_{*}/M_{\odot}$) & $R_{e}$(kpc) & n \\
\hline
357  & 150.23915  & 2.33630  & $H_{\alpha},CO(3-2),CO(1-0)$ & 11.45 &
5.0$\pm$1.2 & 4.4$\pm$0.4  \\
365  & 150.23731 & 2.33815   &
$CO(3-2),CO(1-0)$            & 11.27 &
2.2$\pm$0.2   & 1.6$\pm$0.2 \\
179  & 150.22895 & 2.32975   & $H_{\alpha},CO(3-2),CO(1-0)$ & 11.25 & 
3.7$\pm$0.6   & 0.4$\pm$0.2   \\
372  & 150.23985 & 2.33640   & $H_{\alpha},CO(3-2),CO(1-0)$ & 11.23 &
1.9$\pm$0.3   & 1.9$\pm$0.4  \\
310  & 150.23689 & 2.33579   & $H_{\alpha},CO(3-2),CO(1-0)$ & 11.20 &
1.6$\pm$0.4   & 5.6$\pm$1.0   \\
385  & 150.23864 & 2.33682   & $H_{\alpha},CO(3-2),CO(1-0)$ & 11.18 &
2.2$\pm$0.1   & 1.4$\pm$0.2   \\
270  & 150.23893 & 2.33386   & 
$H_{\alpha}$                 & 11.17 & 
2.7$\pm$0.2   & 2.2$\pm$0.3   \\
622  & 150.23579 & 2.34482   & $H_{\alpha},CO(3-2),CO(1-0)$ & 11.01 & 
1.5$\pm$0.3   & 3.7$\pm$0.6  \\
395  & 150.23693 & 2.33751   &
$CO(3-2),CO(1-0)$            & 11.01 & 
3.4$\pm$0.6   & 6.2$\pm$1.2   \\
636  & 150.21918 & 2.34355   & 
$H_{\alpha}$                 & 10.96 & 
1.1$\pm$0.2   & 2.3$\pm$0.3  \\
326  & 150.23709 & 2.33571   &
$H_{\alpha}$                 & 10.95 & 
2.8$\pm$0.4   & 5.2$\pm$0.8 \\
369  & 150.23744 & 2.33613   & $H_{\alpha},CO(3-2),CO(1-0)$ & 10.92 & 
3.5$\pm$2   & 1.3$\pm$0.2  \\
538  & 150.23641 & 2.34871   &
$CO(3-2),CO(1-0)$            & 10.88 & 
3.2$\pm$0.6   & 3.3$\pm$0.3   \\
710  & 150.23450 & 2.34973   & $H_{\alpha},CO(3-2)$         & 10.63 & 
2.0$\pm$0.2   & 1.4$\pm$0.2   \\
549  & 150.23945 & 2.34803   & $H_{\alpha},CO(3-2),CO(1-0)$ & 10.56 & 
4.1$\pm$0.8   & 3.0$\pm$0.4   \\
375  & 150.23417 & 2.33647   &
$CO(3-2),CO(1-0)$            & 10.37 & 
4.2$\pm$0.5   & 1.3$\pm$0.2   \\
600  & 150.24101 & 2.34603   &
$H_{\alpha}$                 & 9.97 & 
1.8$\pm$0.2   & 1.2$\pm$0.1  \\
566  & 150.23451 & 2.34763   & 
$H_{\alpha}$                 & 9.90 & 
2.2$\pm$0.2   & 1.5$\pm$0.2  \\
386  & 150.25534 & 2.33659   & 
$H_{\alpha}$                 & 9.78 & 
0.8$\pm$0.2   & 1.3$\pm$0.3   \\
88   & 150.24362 & 2.32406   & 
$H_{\alpha}$                 & 9.72 & 
1.5$\pm$0.3   & 5.8$\pm$1.0   \\
679  & 150.24848 & 2.34763   & 
$H_{\alpha}$                 & 9.65 & 
2.8$\pm$0.4   & 1.5$\pm$0.2  \\
16   & 150.23929 & 2.31756   & 
$H_{\alpha}$                 & 9.60 & 
2.4$\pm$0.2   & 1.8$\pm$0.3   \\
574  & 150.24037 & 2.34655   &
$H_{\alpha}$                 & 9.49 & 
0.5$\pm$0.2   & 5.0$\pm$0.9  \\ 
\hline
\end{tabular}  
\end{threeparttable}
}
\end{table*}

\bibliography{main}
\bibliographystyle{aasjournal}

\end{document}